# Mapping a Decade of Avian Influenza Research (2014-2023): A Scientometric Analysis from Web of Science

Muneer Ahmad, *PhD*[1], Undie Felicia Nkatv, *PhD*[2], Amrita Sharma[3], Gorrety Maria Juma[4], Nicholas Kamoga[5] and Julirine Nakanwagi[6]
[1]Chief University Librarian, [2]Ag. Deputy University Librarian, [3]Library Assistant, [4]Deputy University Librarian, [5]Systems Librarian, [6]Librarian
[1,.2,4,5,6]The Iddi Basajjabalaba Memorial Library, Kampala International University, Box 20000, Ggaba Road, Kansanga, Uganda
[3]GDC Ghagwal, Jammu, Jammu and Kashmir, India.
E-mail: [1]muneerbangroo@gmail.com; ahmad.muneer@kiu.ac.ug, [2]felicia.undie@kiu.ac.ug, [3]amusharma11321@gmail.com, [4]goretty.juma@kiu.ac.ug, [5]nicholas.kamoga@kiu.ac.ug, [6]Julirine.nakanwa@kiu.ac.ug

## Abstract
This scientometric study analyzes Avian Influenza research from 2014 to 2023 using bibliographic data from the Web of Science database. We examined publication trends, sources, authorship, collaborative networks, document types, and geographical distribution to gain insights into the global research landscape. Results reveal a steady increase in publications, with high contributions from Chinese and American institutions. Journals such as PLoS One and the Journal of Virology published the highest number of studies, indicating their influence in this field. The most prolific institutions include the Chinese Academy of Sciences and the University of Hong Kong, while the College of Veterinary Medicine at South China Agricultural University emerged as the most productive department. China and the USA lead in publication volume, though developed nations like the United Kingdom and Germany exhibit a higher rate of international collaboration. "Articles" are the most common document type, constituting 84.6% of the total, while "Reviews" account for 7.6%. This study provides a comprehensive view of global trends in Avian Influenza research, emphasizing the need for collaborative efforts across borders.

## Introduction
Avian influenza, caused by influenza A viruses, represents a persistent threat to global health, agriculture, and biodiversity. These viruses primarily circulate among wild aquatic birds, serving as natural reservoirs, but have demonstrated an alarming capacity to infect domestic poultry, mammals, and humans (Morin et al., 2018). The high mutation rate and reassortment potential of avian influenza viruses (AIVs) contribute to their genetic diversity, enabling the emergence of novel strains with varying pathogenicity and host range (Taubenberger & Kash, 2010). Among these, highly pathogenic avian influenza (HPAI) strains, particularly H5N1 and H7N9, have caused severe outbreaks in poultry and sporadic but often fatal human infections, raising concerns about pandemic potential (Yuen & Wong, 2005). The ecological interplay between wild birds, poultry, and humans creates complex transmission dynamics that challenge disease control efforts, necessitating a multidisciplinary approach to understanding and mitigating risks.

The economic and public health impacts of avian influenza are profound. Outbreaks in poultry have led to massive culling operations, trade restrictions, and substantial financial losses, disproportionately affecting low-resource farming communities (Gashaw, 2020). Human infections, though less frequent, carry high mortality rates, particularly with H5N1 and H7N9 subtypes, highlighting the zoonotic threat posed by these viruses (Bui et al., 2016). Moreover, the ability of AIVs to acquire mutations facilitating mammalian adaptation underscores the urgency of surveillance and preparedness (Chauché et al., 2018). Despite advances in diagnostics and vaccine development, gaps persist in predicting spillover events, understanding host-specific immune responses, and implementing effective control measures across diverse ecological and socioeconomic settings.

Research gaps in avian influenza studies are multifaceted. While significant progress has been made in characterizing viral genetics and transmission pathways, the mechanisms driving cross-species transmission remain incompletely understood (Ahmed et al., 2024). Host-pathogen interactions, particularly in wild bird reservoirs, are understudied, limiting predictive models of outbreak emergence (Tuncer et al., 2016). Furthermore, disparities in surveillance capacity between high- and low-income regions hinder global risk assessment, and the effectiveness of vaccination strategies varies widely depending on viral evolution and implementation challenges (Swayne & Kapczynski, 2008). These gaps highlight the need for integrated research that bridges virology, epidemiology, and socioecological systems to inform evidence-based interventions.

Human infections with avian influenza H5N1 viruses present distinct clinical manifestations and pathogenic mechanisms that differentiate them from seasonal influenza strains. The included studies collectively demonstrate that H5N1 infection in humans follows an aggressive clinical course, characterized by rapid progression to severe respiratory distress and multi-organ failure. Studies examining the natural history of human H5N1 disease have identified consistent patterns of high fever, cough, and dyspnea as initial symptoms, frequently progressing to pneumonia and acute respiratory distress syndrome (ARDS) within days of onset (Yuen et al., 1998). The case fatality rate remains alarmingly high, with systemic complications including encephalitis, renal impairment, and hematologic abnormalities contributing to poor outcomes (Peiris et al., 2004).

Comparative analyses of H5N1 isolates from different outbreak periods reveal evolving virulence characteristics. The 2004 Asian isolates demonstrated enhanced pathogenicity in mammalian models compared to earlier strains, with increased lethality observed in ferret studies, (Govorkova et al., 2005; Maines et al., 2005). This phenotypic change correlated with specific genetic markers, including mutations in the hemagglutinin (HA) cleavage site and polymerase complex, which facilitate systemic viral replication. The zoonotic transmission of purely avian viruses to humans without prior adaptation represents a significant departure from conventional influenza epidemiology, challenging existing paradigms about species barriers (Claas et al., 1998).

## Literature review

Bibliometric research has evolved as a vital methodology for evaluating scholarly productivity, impact, and collaboration within higher education institutions. Early studies primarily focused on publication counts and citation frequencies, but later approaches incorporated network analysis to explore collaboration patterns and knowledge dissemination (Hicks et al., 2015). With the advancement of global databases such as Scopus and Web of Science, researchers have conducted extensive assessments that reveal significant disparities in research performance across disciplines and geographic regions (Diem & Wolter, 2013).

Several institutional-level investigations have contributed to understanding research productivity and collaboration trends. Darmadji et al. (2018) examined the Islamic University of Indonesia's scholarly output between 2005 and 2017, noting steady growth in publications but limited collaboration, mostly confined to nearby institutions or alumni networks. They suggested stronger international collaborations and incentives for faculty research engagement. Ahmad and Batcha (2019) analyzed Bharathiar University's publications from 2009 to 2018, documenting 3,440 papers, 38,104 citations, and an h-index of 68. Their use of VOSviewer visualization revealed dynamic patterns in author productivity, institutional collaboration, and journal preferences. Similarly, Kumar and Dora (2012) assessed the Indian Institute of Management Ahmedabad's research contributions, focusing on publication types and citation patterns, while Baskaran (2013) evaluated Alagappa University's performance using Web of Science data. Das et al. (2021) investigated Mizoram University's research output from 2002 to 2018 and found that research articles accounted for 93% of total publications, with 2016 and 2017 as the most productive years. Thapa and Tiwari emerged as leading contributors, with Current Science publishing the largest share of the university's research.

At the national level, Mahala and Singh (2021) examined the scientific output of Indian universities using Web of Science data from 2015 to 2019. Their study retrieved 26,173 documents and revealed that top-performing institutions included the University of Delhi, Banaras Hindu University, Anna University, Jadavpur University, and Punjab University. Multi-authored publications were dominant and tended to receive higher citations, while collaborations were common with countries such as the United States, South Korea, and Germany. Institutional partnerships were especially strong with Anna University, the Indian Institutes of Technology, and the Council of Scientific and Industrial Research (CSIR). Regionally, Ahmad (2022) explored coronary artery disease research within BRICS nations from 1990 to 2019. The findings demonstrated that China led in publication output, while South Africa contributed the least. The study emphasized English-language dominance and the growing trend of collaborative, multi-authored research.

Disciplinary investigations have also highlighted variations in research output across academic ranks and specializations. Matthews (2013) analyzed the productivity of South African physicists between 2009 and 2011 using Web of Science data and departmental records. Professors exhibited higher productivity than lecturers, with overall publication rates comparable to mid-ranked U.S. universities but lower than elite global institutions.

Recent scientometric analyses have focused on African universities, reflecting growing research visibility and collaboration. Ahmad and Nkatv (2025) examined the University of Ibadan's research output from 2014 to 2023, identifying 7,159 publications, 218,572 citations, and an h-index of 75. Their visualization with VOSviewer exposed collaboration strengths and weaknesses, providing evidence-based insights for institutional policy improvements. Likewise, Ahmad and Ubi (2025) analyzed the University of Lagos's research productivity between 2004 and 2023, observing consistent growth, particularly in Health Sciences,

Engineering, and Social Sciences. International collaborations with the United States and the United Kingdom enhanced citation impact, while open-access publishing improved global visibility.

Beyond institutional studies, several scientometric investigations have applied advanced bibliometric tools to specific research areas. Martynov et al. (2020) conducted a scientometric analysis of neuroblastoma research using the Bibliometrix R package, Biblioshiny, VOSviewer, and CiteSpace. Their study analyzed 12,435 documents from 86 countries and found that 12 countries accounted for over 80% of neuroblastoma-related research, collectively receiving 316,017 citations. Similarly, Maghsoudi et al. (2020) examined 50 years of research on sheep red blood cells for monitoring humoral immunity in poultry, analyzing 702 publications from 1968 to 2018 retrieved from Web of Science. The study revealed fluctuations in publication frequency around 1990 and highlighted the link between poultry nutrition and breeding efficiency. Wu et al. (2022) performed a scientometric study on egg yolk immunoglobulin (IgY), analyzing 1,029 papers, 981 journal articles and 48 reviews from Web of Science. China led in publication output, while Dalian University of Technology ranked highest in productivity, and Da Silva's work was the most frequently cited. Wang et al. (2023) conducted a scientometric assessment of betel quid chewing and oral precancerous lesions using Scopus data comprising 1,403 papers, identifying "arecoline" as the most frequent keyword. Zwack et al. (2024) analyzed nearly 13,000 health economics articles from Web of Science, employing co-occurrence and co-citation analyses to classify research into five divisions macroeconomics, microeconomics, outcome valuation, monitoring mechanisms, and appraisal. They found that Europe, North America, and Australia contributed the bulk of literature, with Medicaid expansion identified as a key emerging research stream.

Overall, bibliometric and scientometric research has advanced from simple publication counts to multidimensional analyses that incorporate citation networks, international collaborations, and visualization tools. Across Asia and Africa, institutional and disciplinary studies demonstrate increasing research productivity, though challenges persist in enhancing citation impact, fostering cross-border collaboration, and ensuring equitable research visibility. Tools such as VOSviewer, Biblioshiny, and CiteSpace have been instrumental in mapping research landscapes and supporting evidence-based policymaking in higher education.

## Objectives of the study

The objectives of this study are to:

1. analyze publication trends in Avian Influenza research over the period 2014-2023, including the number of papers published each year and their citation counts;
2. identify the most productive authors and institutions involved in Avian Influenza research and analyze their collaboration patterns;
3. examine the geographical distribution of research on Avian Influenza and identify leading countries and regions contributing to the field;
4. determine the most frequently cited papers, authors, and journals in the Avian Influenza research community;
5. evaluate the impact of specific journals and their contribution to advancing research on Avian Influenza;
6. identify the document type of research in Avian Influenza; and

7. provide recommendations for future research directions based on the analysis of current trends and gaps in Avian Influenza research.

## Methods

This scientometric study on Avian Influenza (2014-2023) utilizes bibliographic data sourced exclusively from the Web of Science, one of the most reputable databases in scientific research. The methodology is structured as follows:

*i. Data collection*:

Source: Bibliographic data was retrieved from the Web of Science database, known for its quality and comprehensive indexing of peer-reviewed literature.

Timeframe: The analysis focused on articles published within the past decade (2014-2023).

Search Criteria: Keywords such as "Avian Flu", "Avian Influenza", "Fowl Plague", and "Influenza, Avian" were used to gather relevant studies.

*ii.* Data processing:

Tools: Data was processed and organized using tools such as RStudio, Histcite, Bibexcel, and MS Excel. These were instrumental in extracting bibliometric indicators such as publication years, citations, authorship, and journal information.

Data Cleaning: Irrelevant or duplicate studies and non-English articles were excluded.

*iii. Bibliometric analysis:*

Publication and Citation Trends: The study analyzed trends in publication volumes and citation patterns, highlighting the influence of key studies over time.

Authorship and Collaboration Networks: Analysis of prolific authors and their collaborations was conducted to identify leading contributors to the field.

Journal Impact: The research assessed the journals that published the most studies on Avian Influenza based on citation counts.

*iv. Visualization:*

Graphical Analysis: Data visualization was carried out using RStudio and VOSviewer to create graphs illustrating trends in publications, citations, author collaborations, and keyword connections.

## Limitations

While this study provides valuable insights into Avian Influenza research, it is subject to several limitations:

i. Database Limitation: The bibliometric data for this study was exclusively sourced from the Web of Science. Other significant databases such as Scopus, PubMed, and Dimensions were not included, which may have led to a narrowed scope of the research landscape. These databases often index different sets of articles, and including them could have provided a more comprehensive view of global research.
ii. Timeframe: The study only considers data from the past 10 years (2014-2023). While this provides a snapshot of recent trends, it may overlook longer-term historical shifts in research on Avian Influenza. Extending the timeframe could have revealed more long-term trends and research development.
iii. Scope of Analysis: The research was focused on a specific set of bibliometric indicators (e.g., citations, authorship, journals, document type), which may have excluded other important dimensions of analysis, such as the qualitative impact of studies or the influence of specific countries or regions.

## Results

### *Main information about the data*

| S.No. | **Table 1: Main Information about Data** | |
|---|---|---|
| 1 | Timespan | 2014:2023 |
| 2 | Sources (Journals, Books, etc) | 1574 |
| 3 | Documents | 10137 |
| 4 | Annual Growth Rate % | -24.68 |
| 5 | Document Average Age | 5.59 |
| 6 | Average citations per doc | 17.8 |
| 7 | References | 234128 |
| | *Document contents* | |
| 8 | Keywords Plus (ID) | 13616 |
| 9 | Author's keywords (DE) | 16304 |
| | *Authors* | |
| 10 | Authors | 32934 |
| 11 | Authors of single-authored docs | 267 |
| | *Authors collaboration* | |
| 12 | Single-authored docs | 377 |
| 13 | Co-authors per Doc | 7.76 |
| 14 | International co-authorships % | 34.52 |

The dataset spans a decade from 2014 to 2023, a period that captures the recent surge of interest in Avian Influenza, including the effects of periodic outbreaks on global health and the response within scientific research. This timeframe is critical as it encompasses significant advances in virology, epidemiology, and policy responses to avian influenza risks. A total of 1,574 sources, including journals, books, and other academic materials, were utilized in this study. This high number of sources indicates a broad multidisciplinary interest and the vast dissemination of knowledge across multiple publication types. The study comprises 10,137 documents, signifying a substantial body of work on Avian Influenza over the past decade. This volume of documents points to the high global research productivity on this topic, likely driven by recurring outbreaks and the associated public health, agricultural, and economic implications. It suggests a robust scientific response to understanding and mitigating Avian Influenza risks. An annual growth rate of -24.68% reflects a decrease in the number of publications over time. The average document age of 5.59 years indicates that the research is relatively recent, with many studies being published in the last few years. This relatively young age reflects the urgency and timeliness of research on Avian Influenza, ensuring that findings remain relevant to current challenges in avian and zoonotic diseases. The average of 17.8 citations per document reflects a high citation impact, suggesting that publications in this area are influential and frequently referenced by the scientific community. With 234,128 references cited across all documents, this figure indicates a rich intertextual landscape within Avian Influenza research, underlining the extent of academic foundation and scholarly engagement with past research. The "Keywords Plus" field includes 13,616 terms, providing insight into the content focus and themes that have emerged as central to Avian Influenza research. A total of 32,934 authors have contributed to this body of research, reflecting substantial collaborative effort and interest across countries and institutions. There are 267 authors of single-authored documents, representing a small subset of the research. This relatively low figure suggests that Avian Influenza research is largely collaborative, likely due to the complexity of the subject, which often requires expertise from multiple fields. There are 377 single-authored documents in the dataset. The modest number of single-authored works aligns with the low number of single-authorship in general, again pointing to the necessity of collaboration in Avian Influenza research. An average of 7.76 co-authors per document reflects a high degree of collaboration, underscoring the multidisciplinary and global nature of Avian Influenza research. The involvement of nearly eight co-authors per publication may imply that researchers from various specialties—such as epidemiology, veterinary science, public health, and virology—are contributing complementary expertise, which is crucial for tackling the multifaceted challenges posed by Avian Influenza. The percentage of international co-authorships at 34.52% signifies a substantial level of international collaboration. This highlights Avian Influenza as a globally relevant issue that transcends borders, with researchers working together across countries to share data, insights, and resources.

### ***Evaluation of the annual output of publications on Avian Influenza research***

**Table 2: Annual distribution of publications and citations**

| S.No. | Year | Records | % | Rank | TLCS | % | TGCS | % |
|---|---|---|---|---|---|---|---|---|
| 1 | 2014 | 1105 | 15.44 | 1 | 8963 | 25.34 | 33644 | 18.65 |
| 2 | 2015 | 1044 | 14.58 | 3 | 7269 | 20.55 | 27885 | 15.45 |
| 3 | 2016 | 990 | 13.83 | 7 | 5325 | 15.05 | 21674 | 12.01 |
| 4 | 2017 | 1078 | 15.06 | 2 | 5186 | 14.66 | 24134 | 13.37 |
| 5 | 2018 | 1008 | 14.08 | 6 | 3077 | 8.70 | 20186 | 11.19 |
| 6 | 2019 | 986 | 13.77 | 8 | 2276 | 6.43 | 17221 | 9.54 |
| 7 | 2020 | 1032 | 14.42 | 4 | 1353 | 3.83 | 16196 | 8.98 |
| 8 | 2021 | 909 | 12.70 | 10 | 945 | 2.67 | 9178 | 5.09 |
| 9 | 2022 | 973 | 13.59 | 9 | 781 | 2.21 | 6710 | 3.72 |
| 10 | 2023 | 1012 | 14.14 | 5 | 196 | 0.55 | 3616 | 2.00 |
| Total | | **10137** | 141.60 | | **35371** | 100.00 | **180444** | 100.00 |

Key: TLCS = Total Local Citation Score; TGCS = Total Global Citation Score

This analysis highlights both the growth in research activity and the citation impact over time, providing insights into publication trends, peaks in research interest, and the longevity of citations. The total publication count over the period is 10,137 documents, with the highest annual contributions from 2014 (15.44%) and 2017 (15.06%). There is a notable drop in publications in 2021 (12.70%) before rising slightly in 2023 (14.14%), indicating fluctuating interest likely due to global events (e.g., COVID-19 may have redirected research focus in recent years). 2014 and 2017 are ranked as the top years for publication volume, with 2014 contributing the highest percentage of records, possibly due to heightened research interest following notable avian influenza outbreaks in the early 2010s. 2014 holds the highest TLCS (8,963; 25.34%), reflecting significant intradataset influence for that year. This high local citation score indicates foundational research that other studies in the dataset heavily referenced.

A declining trend in TLCS from 2014 to 2023 suggests that earlier works (particularly from 2014 to 2017) have had a more lasting impact on subsequent studies. The TLCS value for 2023 is relatively low at 196 (0.55%), indicating that recent publications haven't yet accumulated many citations within this dataset, likely due to limited time for recent research to gain traction. The highest TGCS also belongs to 2014 with 33,644 citations (18.65%), emphasizing the global impact and influence of publications from this year. Recent years (2021–2023) show a drop in TGCS, particularly in 2023 with only 3,616 citations (2.00%), which aligns with the lower TLCS values and underscores the time lag required for newer studies to gain recognition.

The annual distribution of publications and citations reveals a clear trend: Avian Influenza research from 2014–2017 had the highest influence, both locally and globally, suggesting these were formative years for the field. The fluctuating but generally decreasing trend in citation scores from 2018 onwards reflects a combination of research maturity and the emergence of competing health priorities, such as the COVID19 pandemic. However, the overall TGCS and TLCS scores demonstrate the robust, sustained impact of Avian Influenza studies, underscoring their importance in both the scientific community and public health responses. In conclusion, the table 2 showcases how Avian Influenza research has evolved, with early studies continuing to drive scholarly discourse while recent publications await time to accumulate citations. The insights gathered from the table 2 are crucial for understanding the temporal dynamics of Avian Influenza research and the shifting focus within the scientific community.

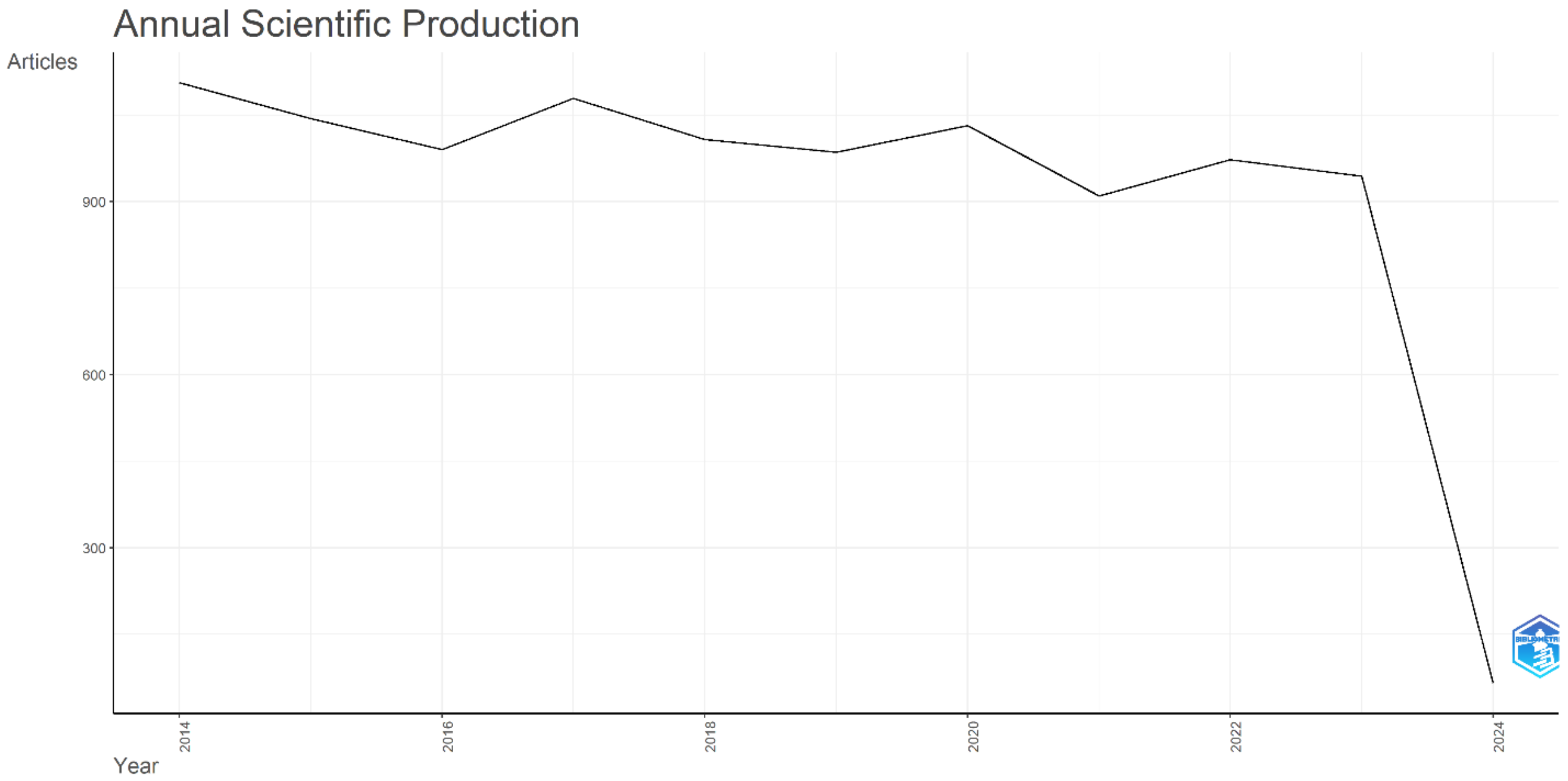

*Figure 1*

## *Analysis of the publication output of top 20 authors on Avian Influenza research*

**Table 3: Publication output of top 20 authors and citation score**

| h-index | Authors | Citation sum within h-core | All citations | All articles |
|---|---|---|---|---|
| 37 | Gao GF | 3833 | 4533 | 82 |
| 35 | Chen HL | 2998 | 4203 | 123 |
| 32 | Fouchier RAM | 4046 | 4843 | 89 |
| 32 | Webby RJ | 2115 | 3132 | 116 |
| 30 | Swayne DE | 2303 | 2944 | 86 |
| 29 | Beer M | 1993 | 2923 | 98 |
| 28 | Kawaoka Y | 2142 | 2555 | 62 |
| 27 | Webster RG | 2732 | 3268 | 78 |
| 27 | Zhang Y | 1884 | 2697 | 107 |
| 26 | Shu YL | 2575 | 3278 | 85 |
| 26 | Bi YH | 2151 | 2728 | 75 |
| 26 | Lee DH | 1916 | 2469 | 94 |
| 26 | Kuiken T | 1944 | 2481 | 68 |
| 25 | Li YB | 1700 | 2107 | 68 |
| 25 | Brown IH | 1797 | 2285 | 72 |
| 24 | Wang DY | 2354 | 2847 | 80 |
| 24 | Krammer F | 2504 | 2716 | 44 |
| 24 | Wang J | 1634 | 1938 | 72 |
| 24 | Chen Y | 1693 | 2167 | 94 |
| 23 | Shi JZ | 1651 | 1887 | 51 |

The Table 3 provides insights into the most influential authors in Avian Influenza research, based on metrics such as: h-index Citation Sum within h-core: Total citations within the h-core, All Citations and All Articles. These metrics reflect both productivity (number of publications) and influence (citations), showcasing the authors whose research has had a significant impact on the field. Gao GF holds the highest h-index (37), with 3,833 citations within the h-core and a total of 4,533 citations across 82 articles. This indicates Gao's extensive influence and productivity, with a solid record of highly cited publications that consistently contribute to the field. Chen HL (h-index of 35) follows closely, with 4,203 total citations across 123 articles. The high article count coupled with a substantial citation total highlights Chen's strong research output and influence. Fouchier RAM stands out for having the highest all citations count (4,843), despite a slightly lower h-index (32), indicating that Fouchier's work may include fewer but highly impactful articles. Fouchier RAM leads with 4,046 citations within the h-core, indicating a significant concentration of citations within the core works. This suggests that Fouchier's research has foundational articles in the field, which are heavily referenced by other researchers.

Authors such as Webster RG and Shu YL also have high h-core citation sums (2,732 and 2,575, respectively), signifying that they have contributed seminal papers that shape Avian Influenza research. Chen HL and Webby RJ are noteworthy for having high article counts (123 and 116, respectively) alongside substantial citation totals. This combination of quantity and quality reflects their sustained contributions to Avian Influenza research. Conversely, authors like Kawaoka Y and Krammer F have lower publication counts (62 and 44, respectively) but have achieved significant impact, as reflected in their citation metrics. This suggests a more focused research output with high relevance. Authors like Webster RG and Kawaoka Y have been major figures in Avian Influenza research over multiple decades, with high citation sums within their h-core. This suggests that their earlier work remains highly relevant and continues to guide newer studies.

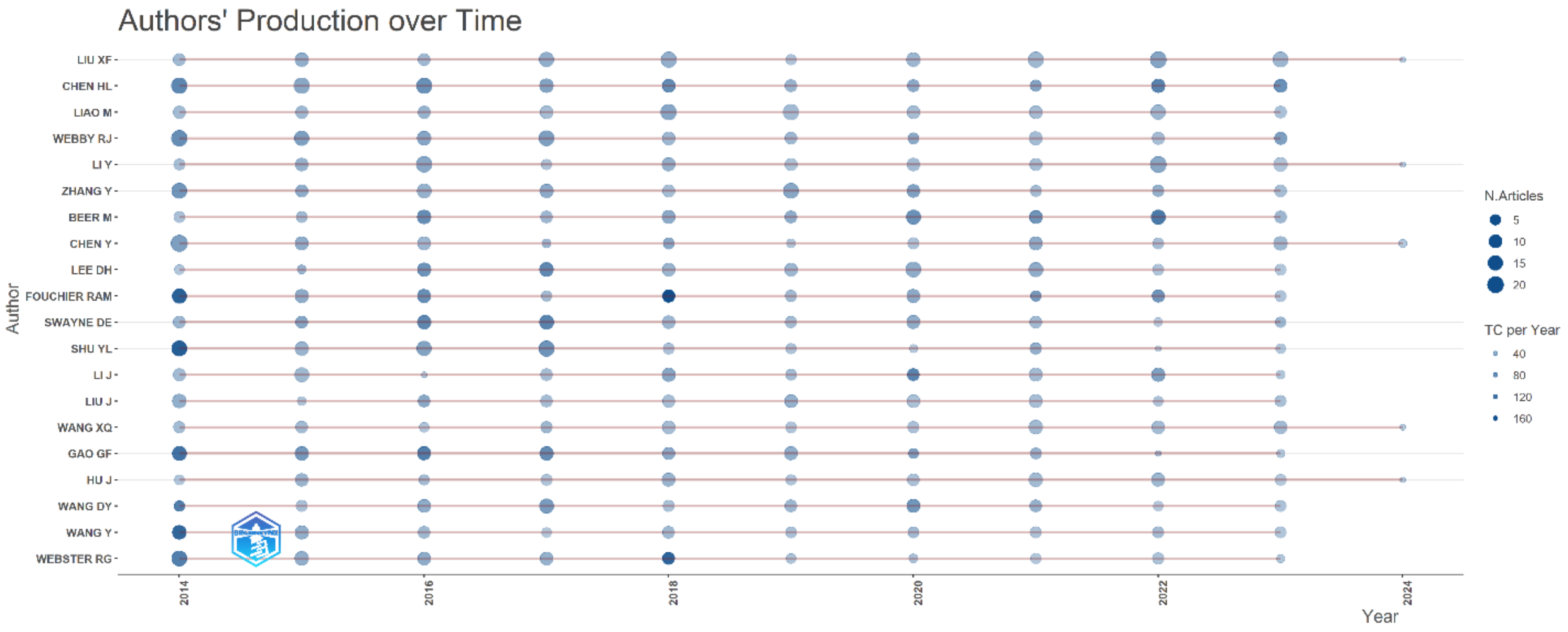

*Figure 2*

The top 20 authors in Avian Influenza research reflect a blend of high productivity, significant influence, and a diverse set of research focuses. Researchers like Gao GF and Chen HL combine high article counts with impactful citations, indicating a prolific and highly influential research presence. On the other hand, authors like Kawaoka Y and Krammer F demonstrate the value of more targeted contributions that achieve considerable citations relative to the number of publications. This analysis shows that Avian Influenza research has been shaped by both a concentrated core of seminal studies and an ongoing expansion of literature by prolific authors. The high h-core citation sums of key authors emphasize the foundational nature of their work, which continues to inform and inspire research in this field.

### *Analysis of source-wise distribution of documents*

**Table 4: Top 20 source wise distribution of documents**

| S.No. | Sources | Documents | h_index | g_index | m_index | TC |
|---|---|---|---|---|---|---|
| 1 | Journal of Virology | 327 | 51 | 71 | 4.64 | 9669 |
| 2 | Emerging Infectious Diseases | 203 | 43 | 63 | 3.91 | 5898 |
| 3 | Scientific Reports | 282 | 38 | 56 | 3.46 | 5936 |
| 4 | Plos One | 359 | 35 | 51 | 3.18 | 6520 |
| 5 | Nature Communications | 45 | 29 | 45 | 2.64 | 2629 |
| 6 | Vaccine | 193 | 29 | 43 | 2.64 | 3393 |
| 7 | Viruses-Basel | 299 | 29 | 50 | 2.64 | 4182 |
| 8 | Veterinary Microbiology | 178 | 28 | 41 | 2.55 | 2700 |
| 9 | Emerging Microbes & Infections | 135 | 27 | 43 | 2.46 | 2779 |
| 10 | Virology | 118 | 26 | 36 | 2.36 | 2098 |
| 11 | Journal of General Virology | 84 | 25 | 36 | 2.27 | 1666 |
| 12 | Plos Pathogens | 65 | 25 | 43 | 2.27 | 1987 |
| 13 | Journal of Infectious Diseases | 71 | 24 | 47 | 2.18 | 2360 |
| 14 | Poultry Science | 147 | 24 | 39 | 2.18 | 2139 |
| 15 | Frontiers in Immunology | 76 | 23 | 42 | 2.30 | 1924 |
| 16 | Proceedings of the National Academy of Sciences of the United States of America | 40 | 23 | 40 | 2.09 | 2875 |
| 17 | Transboundary and Emerging Diseases | 233 | 23 | 30 | 2.09 | 2433 |
| 18 | Avian Diseases | 211 | 22 | 28 | 2.00 | 1970 |
| 19 | Biosensors & Bioelectronics | 32 | 22 | 32 | 2.00 | 1299 |
| 20 | Eurosurveillance | 45 | 22 | 36 | 2.00 | 1330 |

Key: TC= Total Citations

The Table 4 highlights the top 20 journals publishing on Avian Influenza research, ranked by their document output and indexed by various metrics: PLOS ONE published the highest number of articles (359), with an h-index of 35, g-index of 51, and a total of 6,520 citations. Despite the high volume, its citation metrics are moderate, reflecting its broad scope and extensive reach in publishing. Journal of Virology stands out as the top source by h-index (51) and total citations (9,669), across 327 documents. Its high citation impact and comprehensive document output signify its prominence and specialization in virology, including Avian Influenza research. Emerging Infectious Diseases and Scientific Reports also rank high with 203 and 282 documents, respectively, each achieving over 5,000 citations. Their h-indexes of 43 and 38 respectively indicate substantial influence within the field. Nature Communications and PLOS Pathogens have relatively fewer articles (45 and 65, respectively), but boast high citation metrics (h-index of 29 and 25, respectively) and high total citations, indicating that their articles tend to be high-quality and heavily cited. Vaccine, Viruses-Basel, and Veterinary Microbiology also have moderate document counts and high citation totals, suggesting their research on Avian Influenza is both relevant and frequently referenced. Journals like Journal of Virology, Emerging Infectious Diseases, and PLOS ONE demonstrate both high document output and significant citations, showcasing their dual role in promoting both high-quality and frequent research outputs in the field.

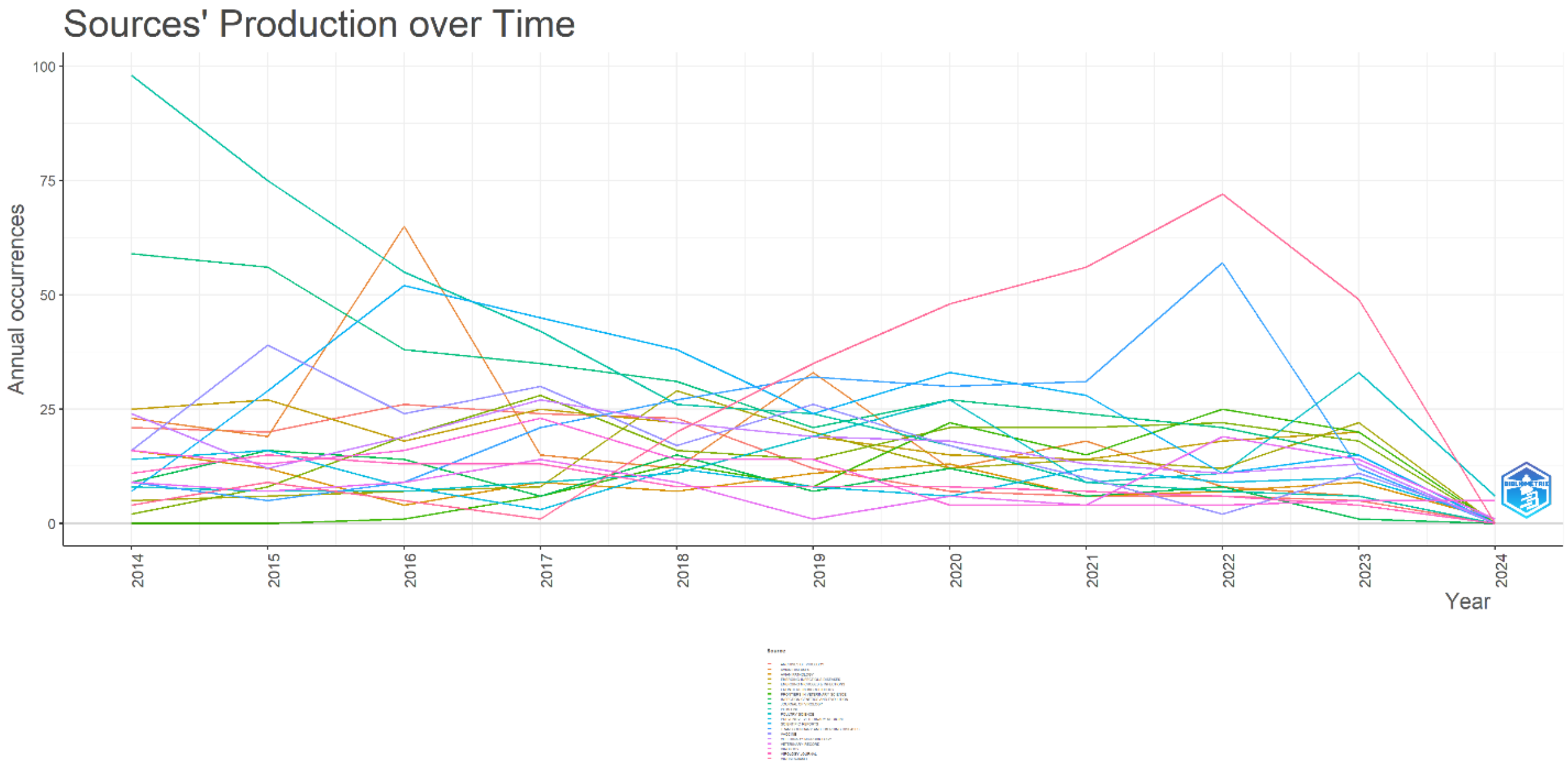

*Figure 3*

Transboundary and emerging diseases and Avian diseases have relatively high document counts (233 and 211, respectively) with reasonable citation metrics, indicating that these journals are valuable platforms for Avian Influenza research and may serve as targeted sources for researchers in this area. The g-index values indicate that journals like Journal of Virology and Emerging Infectious Diseases have numerous articles with high citation counts, emphasizing the presence of landmark studies within these publications. The m-index values for top journals (such as Journal of Virology at 4.64 and Emerging Infectious Diseases at 3.91) suggest these journals have sustained their influence over time, making them reliable sources for impactful research in Avian Influenza.

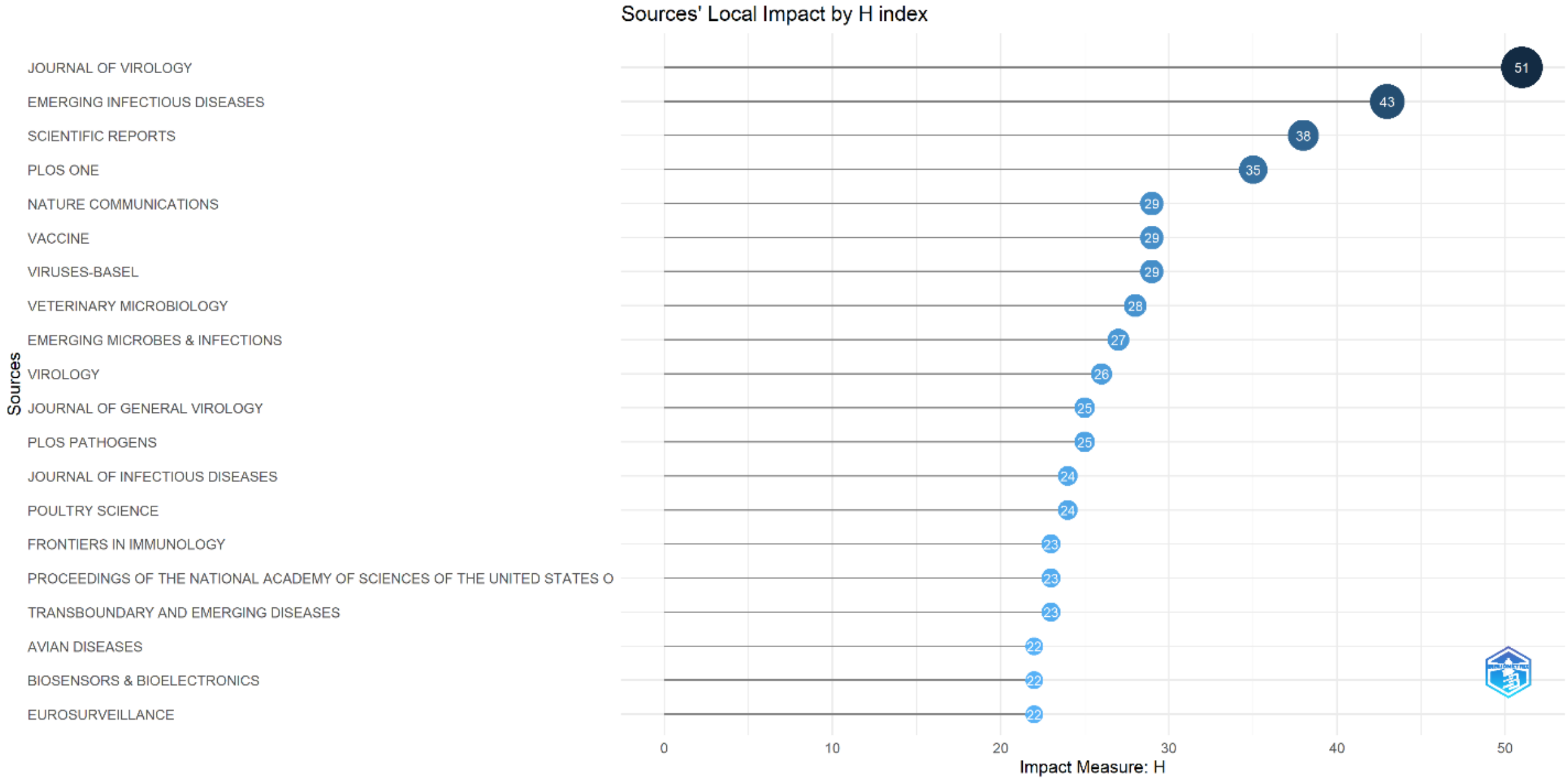

*Figure 4*

The top 20 journals listed here reflect a mix of high-output and high-impact publications in Avian Influenza research. Journals such as Journal of Virology and Emerging Infectious Diseases lead in influence, combining high document counts with substantial citation metrics, and serve as major sources of knowledge. Meanwhile, journals like Nature Communications and PLOS Pathogens highlight the importance of impactful articles that gain significant attention even with lower publication volumes.

The Table 5 ranks the top 20 institutions involved in collaborative research on Avian Influenza. The key metrics include: Chinese Academy of Sciences (CAS) leads with the highest number of records (302), representing 3% of total publications. It has also achieved substantial local (TLCS: 2652) and global citation scores (TGCS: 8779), resulting in an average of 29.07 citations per publication. CAS's output reflects its major influence in Avian Influenza research. Chinese Academy of Agricultural Sciences (CAAS) follows with 297 records, contributing 2.9% of the total output. Despite having fewer citations per publication (ACPP: 18.26) compared to CAS, it is a key player, indicating its strong research focus on agriculture-related aspects of Avian Influenza. Chinese Center for Disease Control and Prevention has one of the highest ACPP scores (42.47), with 137 records. This suggests that its research on Avian Influenza is highly influential within both local and global scientific communities. Erasmus MC (Netherlands) also shows a high ACPP of 42.12 with significant global citations (TGCS: 5855), despite having fewer records (139), highlighting the quality and impact of its publications. The University of Hong Kong is another high-impact institution, with an ACPP of 36.00 and a strong global citation score (TGCS: 8027) across 223 records, showcasing its extensive contributions to Avian Influenza research in Asia. University of Georgia and St. Jude Children's Research Hospital are prominent U.S.-based institutions. While University of Georgia has a moderate ACPP of 20.84, St. Jude achieves a much higher ACPP of 32.99, indicating significant influence.

Centers for Disease Control and Prevention (CDC) also ranks high with strong citation scores (TLCS: 1969, TGCS: 4827), with an ACPP of 26.38, reflecting its role in public health research. University of Oxford and Friedrich Loeffler Institute (Germany) are notable European contributors. Oxford's ACPP of 34.71 and Loeffler's 23.94 reflect high-impact research, showing Europe's significant contributions alongside China and the U.S. Yangzhou University, Zhejiang University, and China Agricultural University represent important collaborative hubs within China. Though their citation impact varies, these institutions underscore China's leading role in Avian Influenza research, focusing on both agriculture and disease control. Institutions like Fudan University also show high ACPP scores (35.82), highlighting impactful research output despite fewer records (117), suggesting a selective focus on high-quality research. US Geological Survey (USGS), US Department of Agriculture (USDA), and Animal and Plant Quarantine Agency (South Korea) have contributed significantly with moderate records but relatively high citation scores, especially USGS and USDA. These agencies focus on the environmental and agricultural impacts of Avian Influenza, essential in controlling the spread within animal populations.

### *Analysis of collaborative ranking of institutions*

**Table 5: Ranking of collaborative institutions**

| S.No. | Institution | Records | % | TLCS | TGCS | ACPP |
|---|---|---|---|---|---|---|
| 1 | Chinese Academy of Science | 302 | 3.00 | 2652 | 8779 | 29.07 |
| 2 | Chinese Academy of Agriculture Science | 297 | 2.90 | 1834 | 5422 | 18.26 |
| 3 | South China Agriculture University | 233 | 2.30 | 1048 | 3730 | 16.01 |
| 4 | University Georgia | 227 | 2.20 | 696 | 4731 | 20.84 |
| 5 | St Jude Childrens Res Hosp | 226 | 2.20 | 1472 | 7455 | 32.99 |
| 6 | University Hong Kong | 223 | 2.20 | 2400 | 8027 | 36.00 |
| 7 | Yangzhou University | 211 | 2.10 | 929 | 2638 | 12.50 |
| 8 | Ctr Dis Control & Prevent | 183 | 1.80 | 1969 | 4827 | 26.38 |
| 9 | Zhejiang University | 166 | 1.60 | 892 | 3046 | 18.35 |
| 10 | ARS | 158 | 1.60 | 1422 | 3478 | 22.01 |
| 11 | China Agriculture University | 152 | 1.50 | 825 | 2835 | 18.65 |
| 12 | Erasmus MC | 139 | 1.40 | 797 | 5855 | 42.12 |
| 13 | Chinese Ctr Disease Control & Prevent | 137 | 1.40 | 2747 | 5819 | 42.47 |
| 14 | Friedrich Loeffler Inst | 137 | 1.40 | 1097 | 3280 | 23.94 |
| 15 | US Geol Survey | 137 | 1.40 | 1234 | 3088 | 22.54 |
| 16 | University Oxford | 123 | 1.20 | 1016 | 4269 | 34.71 |
| 17 | Fudan University | 117 | 1.20 | 709 | 4191 | 35.82 |
| 18 | Minist Agriculture | 110 | 1.10 | 467 | 1623 | 14.75 |
| 19 | USDA | 110 | 1.10 | 1308 | 2461 | 22.37 |
| 20 | Anim & Plant Quarantine Agcy | 106 | 1.00 | 1121 | 1996 | 18.83 |

Key: TLCS = Total Local Citation Score; TGCS = Total Global Citation Score; ACPP= Average Citations Per Paper

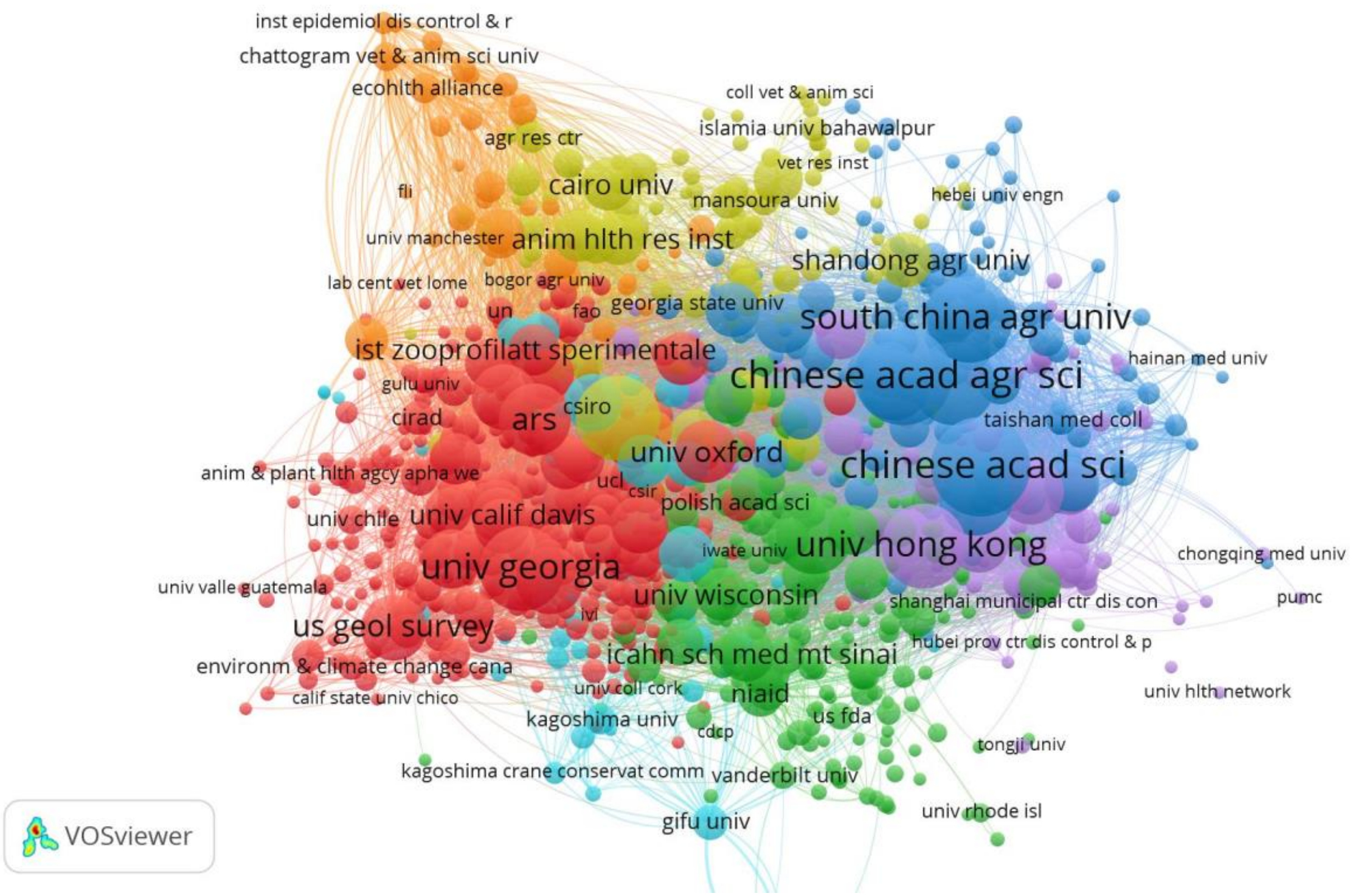

*Figure 5*

The top 20 collaborative institutions in Avian Influenza research reflect a strong international focus, led by Chinese institutions such as Chinese Academy of Sciences and Chinese Academy of Agricultural Sciences, as well as global leaders like University of Hong Kong, St. Jude Children's Research Hospital, and Erasmus MC. High ACPP values for institutions such as the Chinese Center for Disease Control and Prevention and Erasmus MC highlight their impactful contributions to the field. Additionally, significant contributions from U.S.-based institutions and government agencies indicate a well-rounded global effort in addressing Avian Influenza across public health, veterinary sciences, and agricultural sectors. The table 5 emphasizes the extensive collaboration and significant research contributions of institutions worldwide in combating Avian Influenza.

The Table 6 presents the top 20 departments within various institutions that have contributed significantly to Avian Influenza research South China Agricultural University, College of Veterinary Medicine has the most publications (188 records, 1.9% of total output), highlighting its leading role in veterinary-related Avian Influenza research. However, its ACPP is 14.97, indicating moderate impact compared to other departments with fewer publications but higher citation averages. St. Jude Children's Research Hospital, Department of Infectious Diseases ranks second with 165 records and a high ACPP of 31.79, suggesting that its research is not only extensive but also impactful. Icahn School of Medicine at Mount Sinai, Department of Microbiology has the highest ACPP (54.37) despite having fewer records (75), emphasizing the high quality and influence of its publications in the field of Avian

Influenza. University of Hong Kong, Li Ka Shing Faculty of Medicine follows with an ACPP of 56.55 and substantial global citations (TGCS: 3506) across 62 records, indicating that its research is highly regarded globally.

***Analysis of ranking of department-wise distribution***

**Table 6: Ranking of department wise distribution**

| S.No. | Institution with Sub Division | Records | % | TLCS | TGCS | ACPP |
|---|---|---|---|---|---|---|
| 1 | South China Agr Univ, Coll Vet Med | 188 | 1.9 | 943 | 2814 | 14.97 |
| 2 | St Jude Childrens Res Hosp, Dept Infect Dis | 165 | 1.6 | 1056 | 5245 | 31.79 |
| 3 | Chinese Acad Agr Sci, Harbin Vet Res Inst | 138 | 1.4 | 1413 | 3482 | 25.23 |
| 4 | Yangzhou Univ, Coll Vet Med | 111 | 1.1 | 628 | 1515 | 13.65 |
| 5 | Erasmus MC, Dept Virosci | 107 | 1.1 | 533 | 4493 | 41.99 |
| 6 | China Agr Univ, Coll Vet Med | 106 | 1 | 532 | 1880 | 17.74 |
| 7 | Ctr Dis Control & Prevent, Influenza Div | 102 | 1 | 1420 | 3296 | 32.31 |
| 8 | Univ Georgia, Coll Vet Med | 95 | 0.9 | 220 | 1446 | 15.22 |
| 9 | Chinese Acad Sci, Inst Microbiol | 89 | 0.9 | 1286 | 3438 | 38.63 |
| 10 | Yangzhou Univ, Jiangsu Coinnovat Ctr Prevent & Control Important | 84 | 0.8 | 287 | 866 | 10.31 |
| 11 | Icahn Sch Med Mt Sinai, Dept Microbiol | 75 | 0.7 | 406 | 4078 | 54.37 |
| 12 | Chinese Acad Agr Sci, Shanghai Vet Res Inst | 74 | 0.7 | 238 | 1051 | 14.20 |
| 13 | Univ Hong Kong, Sch Publ Hlth | 69 | 0.7 | 600 | 2773 | 40.19 |
| 14 | Friedrich Loeffler Inst, Inst Diagnost Virol | 68 | 0.7 | 500 | 1921 | 28.25 |
| 15 | Konkuk Univ, Coll Vet Med | 67 | 0.7 | 997 | 1822 | 27.19 |
| 16 | Nanjing Agr Univ, Coll Vet Med | 66 | 0.7 | 77 | 625 | 9.47 |
| 17 | Univ Hong Kong, Li Ka Shing Fac Med | 62 | 0.6 | 736 | 3506 | 56.55 |
| 18 | Univ Wisconsin, Sch Vet Med | 59 | 0.6 | 428 | 1934 | 32.78 |
| 19 | Univ Tokyo, Inst Med Sci | 58 | 0.6 | 623 | 2488 | 42.90 |
| 20 | Univ Chinese Acad Sci | 57 | 0.6 | 149 | 1318 | 23.12 |

Key: TLCS = Total Local Citation Score; TGCS = Total Global Citation Score; ACPP= Average Citations Per Paper

Chinese Academy of Agricultural Sciences, Harbin Veterinary Research Institute and Yangzhou University, College of Veterinary Medicine are also notable contributors, reflecting the critical role of veterinary research in understanding and controlling Avian Influenza. Harbin Vet Research Institute has a relatively high ACPP of 25.23, suggesting impactful contributions. Erasmus Medical Center, Department of Viroscience and University of Hong Kong, School of Public Health have high ACPP scores (41.99 and 40.19, respectively), demonstrating the strong impact of specialized departments in virology and public health on Avian Influenza research. Erasmus MC's global citations (TGCS: 4493) further underscore the department's importance in the field. Centers for Disease Control and Prevention, Influenza Division also ranks high with an ACPP of 32.31, underscoring its influence and contribution to Avian Influenza research, particularly in public health. Chinese institutions dominate the table 6, with multiple departments contributing significantly, such as Chinese Academy of Sciences, Institute of Microbiology and Nanjing Agricultural University, College of Veterinary Medicine. This reflects China's prominent role in Avian Influenza research. Other Asian institutions, including University of Tokyo, Institute of Medical Science and Konkuk University, College of Veterinary Medicine in South Korea, also contribute significantly, emphasizing Asia's collaborative efforts in understanding and combating Avian Influenza. Friedrich Loeffler Institute, Institute for Diagnostic Virology shows a high ACPP of 28.25 with 68 records, demonstrating the importance of diagnostic virology in Avian Influenza research. The US institutions like University of Wisconsin, School of Veterinary Medicine also contribute to research in diagnostic virology and epidemiology, showcasing the cross-continental focus on diagnostic and preventive research.

The department-wise analysis highlights the prominent role of veterinary and virology departments across global institutions, particularly in China, the U.S., and Europe. Departments with higher ACPP values, like the Icahn School of Medicine's Department of Microbiology and University of Hong Kong's Li Ka Shing Faculty of Medicine, underscore impactful, high-quality research contributions. Veterinary research departments, such as those at South China Agricultural University and St. Jude Children's Research Hospital, are at the forefront of Avian Influenza research, contributing significantly to both the quantity and impact of publications. The table 6 illustrates the diversity and specialization of departments in addressing Avian Influenza, encompassing fields from veterinary science to microbiology and public health.

### *Distribution of papers by types of documents*

**Table 7: Document type contribution of research**

| S.No. | Document Type | Records | % | TLCS | TGCS |
|---|---|---|---|---|---|
| 1 | Article | 8574 | 84.60 | 28702 | 144517 |
| 2 | Review | 766 | 7.60 | 3205 | 28088 |
| 3 | Letter | 211 | 2.10 | 1912 | 2957 |
| 4 | Editorial Material | 160 | 1.60 | 525 | 1320 |
| 5 | Meeting Abstract | 114 | 1.10 | 26 | 40 |
| 6 | News Item | 93 | 0.90 | 32 | 74 |
| 7 | Article; Proceedings Paper | 73 | 0.70 | 349 | 1019 |
| 8 | Correction | 72 | 0.70 | 7 | 28 |
| 9 | Review; Book Chapter | 19 | 0.20 | 399 | 1209 |
| 10 | Article; Early Access | 17 | 0.20 | 0 | 20 |
| 11 | Article; Retracted Publication | 11 | 0.10 | 81 | 178 |
| 12 | Article; Book Chapter | 7 | 0.10 | 122 | 794 |
| 13 | Retraction | 6 | 0.10 | 0 | 0 |
| 14 | Article; Data Paper | 4 | 0.00 | 0 | 159 |
| 15 | Review; Early Access | 4 | 0.00 | 0 | 9 |
| 16 | Review; Retracted Publication | 2 | 0.00 | 9 | 26 |
| 17 | Biographical-Item | 1 | 0.00 | 0 | 0 |
| 18 | Editorial Material; Early Access | 1 | 0.00 | 0 | 4 |
| 19 | Letter; Retracted Publication | 1 | 0.00 | 2 | 2 |
| 20 | Meeting | 1 | 0.00 | 0 | 0 |
| | **Total** | **10137** | **100.00** | **35371** | **180444** |

Key: TLCS = Total Local Citation Score; TGCS = Total Global Citation Score

Articles represent the majority of the research output, with 8574 records (84.6%), indicating that traditional research articles are the primary format for disseminating Avian Influenza findings. Articles also have the highest TLCS and TGCS values, showing substantial influence within and beyond this dataset. The TGCS for articles is 144,517, reflecting that research articles are widely cited and integral to the field's knowledge base. Review articles make up 7.6% of the total output, with 766 records. Despite a smaller share, reviews have a high TGCS of 28,088 and an average TGCS per review of 36.64, indicating they play a crucial role in summarizing and consolidating Avian Influenza research. This high citation impact shows that review articles are key resources for researchers seeking comprehensive overviews of Avian Influenza developments. Letters account for 2.1% of publications but show high TLCS and TGCS values relative to their quantity. Their TGCS is 2957, indicating that letters, often brief and targeted, are impactful, potentially used for rapid communication of key findings or perspectives in Avian Influenza research. Editorial materials and meeting abstracts have lower record counts (160 and 114, respectively), and their impact, as measured by TLCS and TGCS, is relatively low compared to articles and reviews. However, they contribute to broader scientific discourse, particularly through summarizing key topics and emerging trends presented at conferences. Meeting abstracts, with only 40 global citations, may have limited citation impact, possibly because they represent preliminary findings rather than comprehensive research. Articles categorized as "Proceedings Paper" or "Early Access" have moderate records but generally lower citation scores, indicating that while they contribute to the field, their impact is typically lower than standalone research articles.

Retractions and corrections are also recorded, though they represent a very small proportion of publications, highlighting the efforts to maintain research integrity within the field. Book chapters (both as standalone reviews and articles) are rare but exhibit relatively high TGCS values, suggesting that when Avian Influenza research is published in book form, it remains useful to researchers, possibly due to the in-depth analysis often provided in book chapters. Data papers, although minimal in number, have some TGCS, indicating the value of shared datasets in supporting other researchers. There are 11 retracted articles and 6 standalone retractions, making up less than 0.1% of the total publications. Their presence reflects the field's quality control practices. The citations for retracted papers are low, which could indicate a reduction in impact post-retraction.

The majority of Avian Influenza research is shared through articles and review papers, which are both highly cited and influential in the field. Review articles, though fewer in number, play a critical role by synthesizing existing research. Letters and editorials, while brief, provide impactful contributions, and meeting abstracts showcase ongoing research discussions. Book chapters, although less frequent, offer valuable in-depth analyses. Retractions and corrections, though minor, emphasize a commitment to research quality and integrity. This document type analysis shows that while research articles dominate, a variety of formats contribute to knowledge dissemination in Avian Influenza studies.

***Most relevant countries by corresponding authors***

**Table 8: Most relevant countries by corresponding authors**

| S.No. | Country | Articles | Articles % | SCP | MCP | MCP % |
|---|---|---|---|---|---|---|
| 1 | China | 2896 | 28.60 | 2272 | 624 | 21.50 |
| 2 | USA | 1856 | 18.30 | 1271 | 585 | 31.50 |
| 3 | Korea | 439 | 4.30 | 312 | 127 | 28.90 |
| 4 | United Kingdom | 436 | 4.30 | 232 | 204 | 46.80 |
| 5 | Japan | 366 | 3.60 | 211 | 155 | 42.30 |
| 6 | Germany | 364 | 3.60 | 190 | 174 | 47.80 |
| 7 | Australia | 304 | 3.00 | 127 | 177 | 58.20 |
| 8 | Canada | 268 | 2.60 | 153 | 115 | 42.90 |
| 9 | France | 232 | 2.30 | 119 | 113 | 48.70 |
| 10 | Netherlands | 211 | 2.10 | 118 | 93 | 44.10 |
| 11 | Egypt | 196 | 1.90 | 85 | 111 | 56.60 |
| 12 | India | 175 | 1.70 | 144 | 31 | 17.70 |
| 13 | Italy | 154 | 1.50 | 79 | 75 | 48.70 |
| 14 | Iran | 134 | 1.30 | 111 | 23 | 17.20 |
| 15 | Spain | 127 | 1.30 | 72 | 55 | 43.30 |
| 16 | Pakistan | 117 | 1.20 | 79 | 38 | 32.50 |
| 17 | Poland | 115 | 1.10 | 90 | 25 | 21.70 |
| 18 | Thailand | 110 | 1.10 | 65 | 45 | 40.90 |
| 19 | Russia | 93 | 0.90 | 53 | 40 | 43.00 |
| 20 | Brazil | 88 | 0.90 | 59 | 29 | 33.00 |

Key: SCP=Single Country Publications; MCP=Multiple Country Publications

The Table 8 ranks countries by the number of Avian Influenza research articles attributed to corresponding authors from each country. China ranks first, contributing 2896 articles (28.6%). Most of these (2272) are single-country publications (SCP), while 21.5% are multi-country collaborations (MCP), suggesting a strong domestic research output with moderate international collaboration. The USA follows with 1856 articles (18.3%). While the USA has fewer publications than China, 31.5% of these involve multi-country collaboration, indicating a greater tendency for international partnerships. The United

Kingdom, Germany, and Australia have some of the highest MCP percentages (46.8%, 47.8%, and 58.2%, respectively), showing a strong inclination toward collaborative research. This trend suggests that Avian Influenza research in these countries is significantly influenced by international partnerships, potentially enhancing the diversity of research perspectives. Korea ranks third with 439 articles (4.3%), and 28.9% of these are MCPs, reflecting active engagement in global research networks. Japan also has a high percentage of MCPs (42.3%), underscoring its commitment to collaborative research in Avian Influenza

.

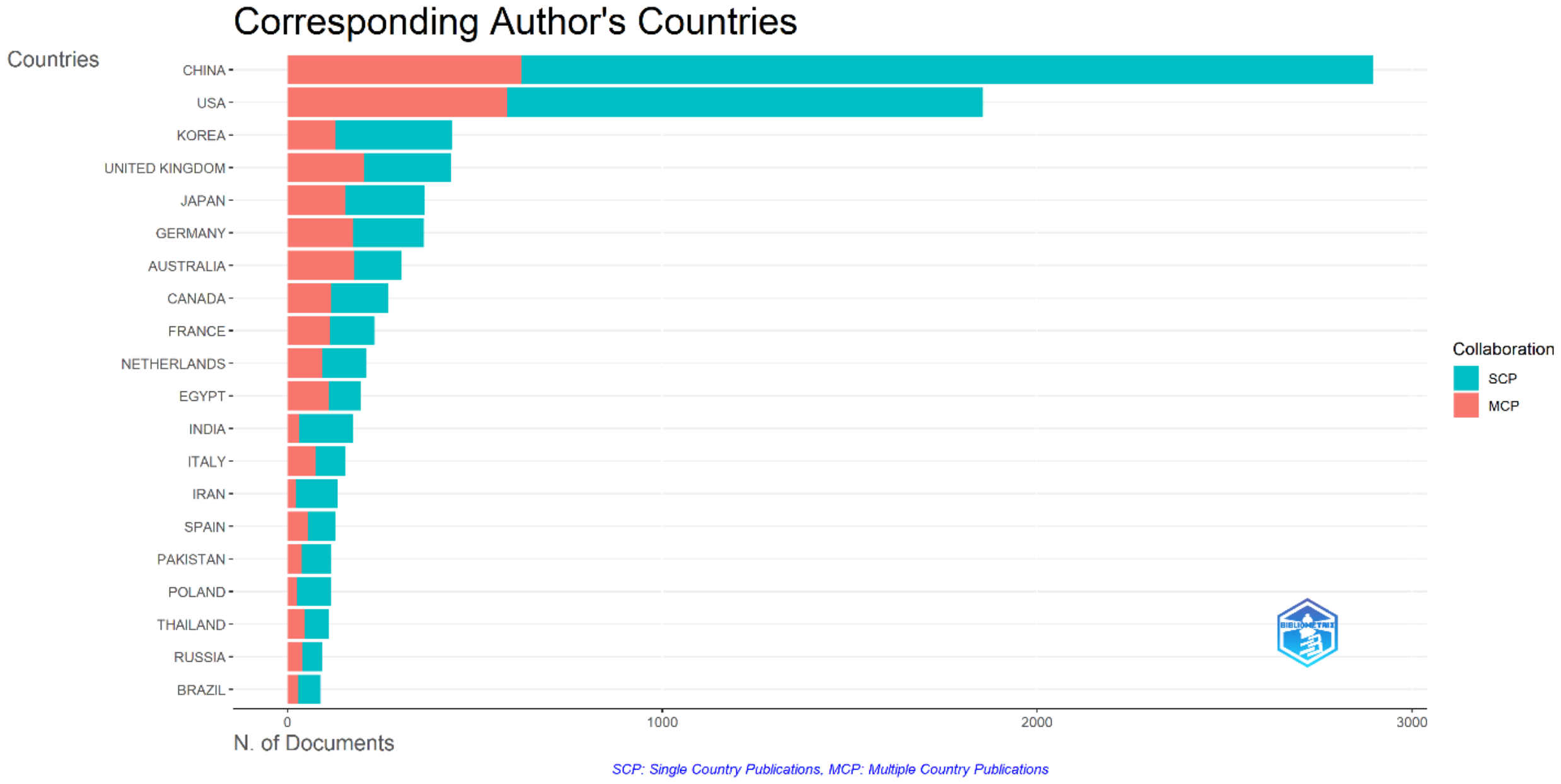

*Figure 6*

India and Iran, though contributing fewer articles (175 and 134, respectively), show lower MCP percentages (17.7% and 17.2%), suggesting that their research efforts are primarily national rather than international. Germany, France, and Italy all show high MCP percentages (47.8%, 48.7%, and 48.7%, respectively), reflecting their extensive engagement in international research networks. Netherlands and Spain also have high MCP percentages, contributing to a collaborative European research environment. Egypt and Pakistan have notable MCP percentages (56.6% and 32.5%, respectively). Egypt's high MCP percentage suggests that much of its Avian Influenza research is done in collaboration with other countries, likely due to shared concerns about the disease's impact. Brazil and Thailand also have meaningful MCP percentages, showing collaborative engagement from emerging economies in Avian Influenza research. Poland and Russia contribute a moderate number of articles, with MCP percentages of 21.7% and 43%, respectively, indicating that Russia is more active in collaborative research than Poland.

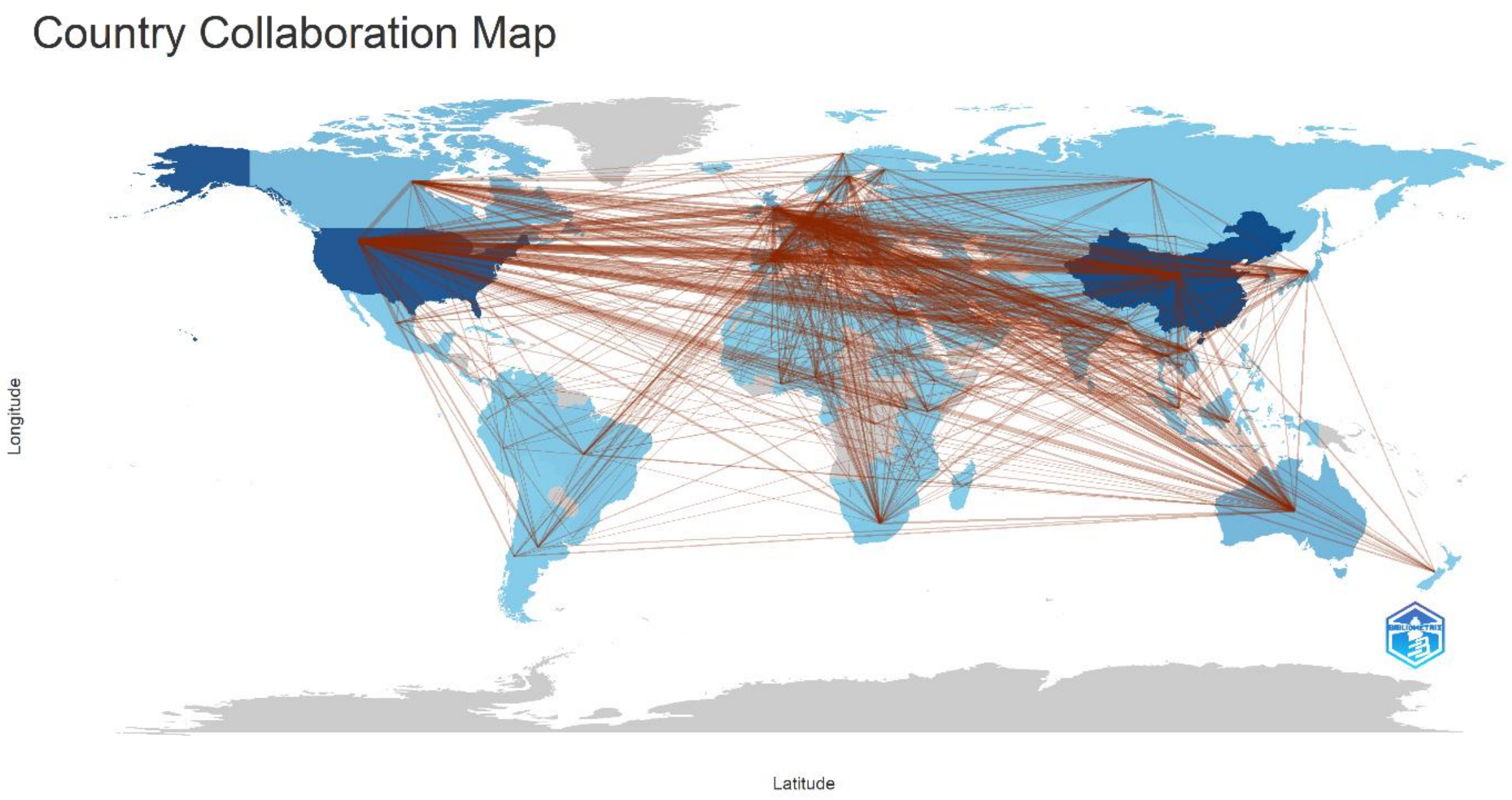

*Figure 7*

China and the USA lead in publication volume, while European and developed nations generally show high levels of international collaboration. Countries like Egypt and Australia have high MCP percentages, indicating substantial collaborative efforts. In contrast, nations like India and Iran contribute primarily through single-country studies. The varied levels of international collaboration highlight a globally diverse research landscape in Avian Influenza studies, with both developed and emerging economies playing significant roles in advancing knowledge in this field.

## Discussion

The findings highlight the concentrated research efforts on Avian Influenza in specific countries and institutions, with China and the USA leading both in quantity and impact. While Chinese institutions like the Chinese Academy of Sciences and the South China Agricultural University prioritize high-volume output, Western institutions, particularly in the United Kingdom and Germany, show strong multi-country publication rates, indicating a focus on collaborative, interdisciplinary research. Departments specializing in veterinary medicine, microbiology, and infectious diseases are major contributors, reflecting the health and agricultural implications of Avian Influenza. The dominance of "Articles" and "Reviews" among document types underscores the prevalence of empirical studies and review-based synthesis in this field.

## Conclusion

This scientometric analysis demonstrates that Avian Influenza research is robust but geographically concentrated. While China and the USA dominate in volume, developed countries, especially in Europe, emphasize collaborative approaches. Veterinary and medical sciences play a central role in the research output, highlighting the disease's impact on both human and animal health. These findings underscore the need for sustained global collaboration and cross-disciplinary research to address Avian Influenza's complex challenges. This research is critical as Avian Influenza poses significant threats to global health, agriculture, and economies. By mapping the landscape of Avian Influenza research, this study identifies leading institutions, countries, and collaborative patterns, enabling policymakers, researchers, and funding bodies to strategize resource allocation and promote collaborations. Understanding publication trends also helps scientists to pinpoint research gaps and target areas requiring further exploration. Through these insights, this study supports the development of effective preventive and response strategies for Avian Influenza on a global scale.

**Recommendations**

To strengthen the global Avian Influenza research landscape, there is a need to enhance international and interdisciplinary collaborations, particularly linking high-output countries with underrepresented regions. Funding agencies should support research expansion in emerging areas while encouraging integrative work across veterinary, medical, and public health domains. Greater attention should be given to understudied aspects such as emerging strains, surveillance innovations, and early-warning systems. Improving global data-sharing mechanisms will also help accelerate scientific progress and strengthen preparedness efforts.

**Disclosure statement**

There is no conflict of interest reported by the authors

.

**References**

Ahmad, M. (2022). *Mapping research productivity of BRICS countries with special reference to coronary artery disease (CAD): A scientometric study* [Doctoral dissertation, Annamalai University]. http://hdl.handle.net/10603/460776

Ahmad, M., & Batcha, M. S. (2019). Scholarly communications of Bharathiar University on Web of Science in global perspective: A scientometric assessment. *Research Journal of Library and Information Science, 3*(3), 22–29.

Ahmad, M., & Nkatv, U. F. (2025). Measuring the research output and performance of the University of Ibadan from 2014 to 2023: A scientometric analysis. *Nigerian Libraries, 59*(1), 1–16. https://doi.org/10.61955/HFYDJH

Ahmad, M., & Ubi, S. I. (2025). Two decades of research at the University of Lagos (2004–2023): A scientometric analysis of productivity, collaboration, and impact. *Research & Reviews: Journal of Statistics, 14*(1), 21–37. https://journals.stmjournals.com/rrjost/article=2025/view=208027

Ahmed, M., M., N. M., Aitor, N., S., B. R., P., S. J., Adolfo, G.-S., & Luis, M.-S. (2024). Avian influenza A (H5N1) virus in dairy cattle: origin, evolution, and cross-species transmission. *MBio*, *15*(12), e02542-24. https://doi.org/10.1128/mbio.02542-24

Baskaran, C. (2013). Research productivity of Alagappa University during 1999–2011: A bibliometric study. *DESIDOC Journal of Library and Information Technology, 33*(3), 236–242. https://doi.org/10.14429/djlit.33.3.4609

Bui, C., Bethmont, A., Chughtai, A. A., Gardner, L., Sarkar, S., Hassan, S., Seale, H., & MacIntyre, C. R. (2016). A Systematic Review of the Comparative Epidemiology of Avian and Human Influenza A H5N1 and H7N9 - Lessons and Unanswered Questions. *Transboundary and Emerging Diseases*, *63*(6), 602–620. https://doi.org/10.1111/tbed.12327

Chauché, C., Nogales, A., Zhu, H., Goldfarb, D., Ahmad, S. A. I., Gu, Q., Parrish, C. R., Martínez-Sobrido, L., Marshall, J. G., & Murcia, P. R. (2018). Mammalian Adaptation of an Avian Influenza A Virus Involves Stepwise Changes in NS1. *Journal of Virology*, *92*(5), 10.1128/jvi.01875-17. https://doi.org/10.1128/jvi.01875-17

Claas, E. C., Osterhaus, A. D., van Beek, R., De Jong, J. C., Rimmelzwaan, G. F., Senne, D. A., Krauss, S., Shortridge, K. F., & Webster, R. G. (1998). Human influenza A H5N1 virus related to a highly pathogenic

avian influenza virus. *Lancet (London, England)*, *351*(9101), 472–477. https://doi.org/10.1016/S0140-6736(97)11212-0

Darmadji, A., Prasojo, L. D., Riyanto, Y., Kusumaningrum, F. A., & Andriansyah, Y. (2018). Publications of Islamic University of Indonesia in Scopus database: A bibliometric assessment. *COLLNET Journal of Scientometrics and Information Management, 12*(1), 109–131. https://doi.org/10.1080/09737766.2017.1400754

Das, S., Kumar, S., Manoj, Y., Verma, K., Das, M. S., Yadav, S. K., & Manoj Kumar Verma, D. (2021). Research productivity of Mizoram University, Aizawl during 2002–2018: A bibliometric analysis. *Journal of Indian Library Association, 56*(3), 1–11. https://www.ilaindia.net/jila/index.php/jila/article/view/382

Diem, A., & Wolter, S. C. (2013). The use of bibliometrics to measure research performance in education sciences. *Research in Higher Education, 54*(1), 86–114. https://doi.org/10.1007/s11162-012-9264-5

Gashaw, M. (2020). A Review on Avian Infl uenza and its Economic and Public Health Impact. *Int J Vet Sci Technol*, *4*(1), 15–027. www.scireslit.com

Govorkova, E. A., Rehg, J. E., Krauss, S., Yen, H.-L., Guan, Y., Peiris, M., Nguyen, T. D., Hanh, T. H., Puthavathana, P., Long, H. T., Buranathai, C., Lim, W., Webster, R. G., & Hoffmann, E. (2005). Lethality to ferrets of H5N1 influenza viruses isolated from humans and poultry in 2004. *Journal of Virology*, *79*(4), 2191–2198. https://doi.org/10.1128/JVI.79.4.2191-2198.2005

Hicks, D., Wouters, P., Waltman, L., de Rijcke, S., & Rafols, I. (2015). Bibliometrics: The Leiden manifesto for research metrics. *Nature, 520*(7548), 429–431. https://doi.org/10.1038/520429a

Kumar, A. H., & Dora, M. (2012). Research productivity in a management institute: An analysis of research performance of Indian Institute of Management Ahmedabad during 1999–2010. *DESIDOC Journal of Library and Information Technology, 32*(4), 365–372. https://doi.org/10.14429/djlit.32.4.2533

Mahala, A., & Singh, R. (2021). Research output of Indian universities in sciences (2015–2019): A scientometric analysis. *Library Hi Tech, 39*(4), 984–1000. https://doi. \org/10.1108/LHT-09-2020-0224

Maines, T. R., Lu, X. H., Erb, S. M., Edwards, L., Guarner, J., Greer, P. W., Nguyen, D. C., Szretter, K. J., Chen, L.-M., Thawatsupha, P., Chittaganpitch, M., Waicharoen, S., Nguyen, D. T., Nguyen, T., Nguyen, H. H. T., Kim, J.-H., Hoang, L. T., Kang, C., Phuong, L. S., … Tumpey, T. M. (2005). Avian influenza (H5N1) viruses isolated from humans in Asia in 2004 exhibit increased virulence in mammals. *Journal of Virology*, *79*(18), 11788–11800. https://doi.org/10.1128/JVI.79.18.11788-11800.2005

Matthews, A. P. (2013). Physics publication productivity in South African universities. *Scientometrics, 95*(1), 69–86. https://doi.org/10.1007/s11192-012-0842-2

Maghsoudi, A., Vaziri, E., Feizabadi, M., & Mehri, M. (2020). Fifty years of sheep red blood cells to monitor humoral immunity in poultry: A scientometric evaluation. *Poultry Science, 99*(10), 4758–4768. https://doi.org/10.1016/j.psg.2020.06.058

Martynov, I., Frysch, J. K., & Schoenberger, J. (2020). A scientometric analysis of neuroblastoma research. *BMC Cancer, 20*(1), 1–13. https://doi.org/10.1186/s12885-020-06974-3

Morin, C. W., Stoner-Duncan, B., Winker, K., Scotch, M., Hess, J. J., Meschke, J. S., Ebi, K. L., & Rabinowitz, P. M. (2018). Avian influenza virus ecology and evolution through a climatic lens. *Environment International*, *119*, 241–249. https://doi.org/https://doi.org/10.1016/j.envint.2018.06.018

Peiris, J. S. M., Yu, W. C., Leung, C. W., Cheung, C. Y., Ng, W. F., Nicholls, J. M., Ng, T. K., Chan, K. H., Lai, S. T., Lim, W. L., Yuen, K. Y., & Guan, Y. (2004). Re-emergence of fatal human influenza A subtype H5N1 disease. *The Lancet*, *363*(9409), 617–619. https://doi.org/10.1016/S0140-6736(04)15595-5

Swayne, D. E., & Kapczynski, D. (2008). Strategies and challenges for eliciting immunity against avian influenza virus in birds. *Immunological Reviews*, *225*(1), 314–331. https://doi.org/10.1111/j.1600-065X.2008.00668.x

Taubenberger, J. K., & Kash, J. C. (2010). Influenza Virus Evolution, Host Adaptation, and Pandemic Formation. *Cell Host & Microbe*, *7*(6), 440–451. https://doi.org/10.1016/j.chom.2010.05.009

Tuncer, N., Torres, J., Martcheva, M., Barfield, M., & Holt, R. D. (2016). Dynamics of low and high pathogenic avian influenza in wild and domestic bird populations. *Journal of Biological Dynamics*, *10*(1), 104–139. https://doi.org/10.1080/17513758.2015.1111449

Wang, G., Yu, Z., Ji, T., Shi, L., & Liu, W. (2023). A scientometric study of betel quid chewing and oral cancer and precancerous lesions with distinct regional characteristics. *Journal of Dental Sciences, 18*(3), 1378–1383. https://doi.org/10.1016/j.ds.2023.03.007

Wu, R., Yakhkeshi, S., & Zhang, X. (2022). Scientometric analysis and perspective of IgY technology study. *Poultry Science, 101*(4), 101713. https://doi.org/10.1016/j.psj.2022.101713

Yuen, K. Y., Chan, P. K. S., Peiris, M., Tsang, D. N. C., Que, T. L., Shortridge, K. F., Cheung, P. T., To, W. K., Ho, E. T. F., Sung, R., & Cheng, A. F. B. (1998). Clinical features and rapid viral diagnosis of human disease associated with avian influenza A H5N1 virus. *The Lancet*, *351*(9101), 467–471. https://doi.org/10.1016/S0140-6736(98)01182-9

Yuen, K. Y., & Wong, S. S. Y. (2005). Human infection by avian influenza A H5N1. *Hong Kong Medical Journal = Xianggang Yi Xue Za Zhi*, *11*(3), 189–199.

Zwack, C. C., Haghani, M., & Grob, E. W. B. (2024). Research trends in contemporary health economics: A scientometric analysis on collective content of specialty journals. *Health Economics Review, 14*(6). https://doi.org/10.1186/s13561-023-00471-6